\documentclass[12pt]{article}   
\usepackage{latexsym} 
\usepackage{mathptm}
\usepackage{amsmath}
\usepackage{amssymb}
\usepackage{graphicx}

\def\beq{\begin{equation}}
\def\eeq{\end{equation}}
\def\beqn{\begin{eqnarray}}
\def\eeqn{\end{eqnarray}}

\begin{document}
 
\title{Comments on Nonlocality in Deformed Special Relativity, in reply to arXiv:1004.0664 by Lee 
Smolin and arXiv:1004.0575 by Jacob et al.}
\author{Sabine Hossenfelder \thanks{hossi@nordita.org}\\
{\footnotesize{\sl NORDITA, Roslagstullsbacken 23, 106 91 Stockholm, Sweden}}}
\date{}
\maketitle
 
\vspace*{-1cm}
\begin{abstract}
It was previously shown that models with deformations of special relativity that
have an energy-dependent yet observer-independent speed of light suffer
from nonlocal effects that are in conflict with observation to very high precision. In a
recent paper it has been proposed that these paradoxa arise only in the classical limit and
can be prevented by an ad-hoc introduction of a quantum uncertainty that would
serve to hide the nonlocality. We will show here that the proposed ansatz for this
resolution is inconsistent with observer-independence and, when corrected, is in 
agreement with the earlier argument that revealed the troublesome nonlocality. 
We further offer an alternative derivation for the energy-dependent speed of light
in the model used.
\end{abstract}

\section{What the argument is about}

It has been claimed that deformations of special relativity ({\sc DSR}) \cite{DSR} make it
possible to introduce an energy-dependent speed of light in position
space while preserving observer-independence. Or, to be exact, since there is no 
(agreed upon) formulation of {\sc DSR} in position space the energy-dependent 
speed of light $\tilde c(E)$ that one finds in these models in momentum space 
has been used as a speed in position
space \cite{AmelinoCamelia:2000ge}. In particular it has been used to make predictions for a possible time-delay
in the arrival time of high energetic photons from distant gamma ray bursts that
might be on the edge of detection. As an observable prediction of quantum 
gravitational effects, this has received a lot of attention \cite{Science,Nature,Albert:2007qk,AmelinoCamelia:2009pg}.

It goes without saying that the lacking derivation of the photon's propagation in
position space is a severe shortcoming of the model and raises strong doubts about
the use of the prediction to begin with. Moreover, 
in \cite{Hossenfelder:2009mu} it
has been argued that even absent a derivation of the propagation in position
space, it can be concluded that an energy-dependent and observer-independent speed
of light of the sort that leads to the predictions made is in conflict with
observations already, at least to first order in energy over Planck mass, $E/m_{\rm p}$. 

It should be emphasized that in \cite{Hossenfelder:2009mu} it has not been
shown that {\sc DSR} does have an energy-dependent speed of light in position 
space\footnote{It has been claimed in \cite{KowalskiGlikman:2001px} that the
speed of massless particles is constant in {\sc DSR}.}.
The logic of the argument presented in \cite{Hossenfelder:2009mu} is instead to
assume that there is indeed such an energy-dependent speed of light that would lead to an
observable time-delay in high energetic gamma ray bursts, and that this effect
would also be observer-independent. Then it can be shown that this leads to
strongly nonlocal effects that would long have been observed. 
This means in return, since we have not observed such effects, {\sc DSR} either does
not have an energy-dependent speed of light in position space, or it does
in which case it is not compatible with reality. Either way one can conclude
that {\sc DSR} cannot give rise to observable time-delays in gamma ray bursts
in contrast to the predictions made.

Of course this conclusion depends on the assumptions made, the details 
can be found in \cite{Hossenfelder:2009mu}. It should also be mentioned 
that some problems with locality in {\sc DSR} had been pointed out previously
by Unruh and Sch\"utzhold \cite{Schutzhold:2003yp} but the nonlocality 
had not been quantified. It had been claimed by Amelino-Camelia
that such nonlocal effects ``can be safely neglected'' \cite{AmelinoCamelia:2002vy}. 

The reason for the arising nonlocality is, in short, as follows. In special
relativity there is exactly one speed that is observer-independent, that
is the usual, constant, speed of light $c$. In {\sc DSR} now photons with different
energies travel with different speeds, depending on their energy.
 If this speed of light $\tilde c(E)$ is observer-independent, an observer
who measures the energy $E'$ has to find $\tilde c(E')$ for its speed. It is easy
to see that this is not possible with usual Lorentz-transformations. To
achieve the invariance of the speed of light one has to use modified
Lorentz-transformations. These transformations then have to depend on the energy
of the particle. 

Leaving aside for a moment the interpretational problems with a change
of reference frame depending on an energy (which?), this causes another
problem: Since the crossing of worldlines defines events
in space-time, using these modified, energy-dependent, transformations 
changes the location in which two world-lines meet relative to the
special relativity case. For two worldlines
this might not seem too worrisome since there is no absolute meaning to
position anyway and (in 1+1 dimensions) the two lines will generically meet in some point. But
if one considers three worldlines that in one frame meet in one point,
then the use of the modified transformation will generically 
cause this point to split up in three different points. Or rather, there
is then no such thing as ``one point'' -- it's an ill-defined concept. This
is a consequence of the three worldlines transforming differently now.
In the special relativity case, three lines that meet in one point in
one reference frame will meet in one point in all reference frames.

How much the point splits up depends on the distance the particles have
traveled. This is because the modified transformations change, diagrammatically
speaking, the angle of the worldline relative to the usual Lorentz-transformation, 
thus the total change adds up over the distance. This is the same reason why the 
effect has been claimed to be observable in the first place. It can
be shown that the splitting of the point for realistic distances and
particle energies can be as large as $\approx 1$km just by changing from the Earth restframe
to a satellite in Earth orbit. One might have expected
some sort of deviation from the usual space-time picture, but it should
not happen on such macroscopic scales.

This is the simple case of the locality problem that has 
in \cite{Hossenfelder:2009mu} been referred to as Version 2.0. In
Version 2.1 it has further been considered the option that {\sc DSR}
in position space also causes a spread of a (usually dispersionless) 
photon wave-packet. This spread has then to be compared to the splitting
of the points to see whether it can hide the nonlocality. It has
been shown that while such a spread vastly improves the problem, it is still 
not possible to hide the nonlocality in all reference frames,
and that even moderate boosts (up to $\gamma\approx 20$) are sufficient
to reveal the nonlocality again.

\section{The proposed solution}

In a recent paper by Lee Smolin \cite{Smolin:2010xa} it has now been proposed an ansatz 
for quantum uncertainties that could potentially address the problem of
nonlocality in {\sc DSR}. The idea is that the nonlocality, the splitting-up of the point, is entirely hidden by quantum position 
uncertainty. What one needs is that the uncertainty in a particle's 
position is so large that the probability of it being at any of
the three split-up points is about the same in all restframes. Then, for what
the particle is concerned, the three points could not be resolved and
it would be appropriate to understand them as one large point, removing
the inconsistency.

In \cite{Smolin:2010xa} it has not been examined the three-particle 
case discussed in \cite{Hossenfelder:2009mu} and it was left open for
future studies whether the proposed ansatz would indeed solve the problem. 
Before we look into the proposal of \cite{Smolin:2010xa}, let us note that even if
the attempt to hide the nonlocality by additional uncertainty was successful, it would
still mean that previously made predictions for gamma ray bursts were
wrong (as also pointed out in \cite{Smolin:2010xa}). The effect would then not present itself in an either earlier
or later arrival of highly energetic photons, but instead in a
stochastic spread over arrival times where the width of the spread
depends on the energy. 

The ansatz made in \cite{Smolin:2010xa}, see Eqs.(44,45), is
\beqn
\Delta x_{AliceQ} \approx l_{\rm p} |a| |p| ~,~ \Delta T_{AliceQ} \approx l_{\rm p} |a| |E|  \quad, 
\label{ansatz}
\eeqn
where we have set $\hbar =1$ and $l_{\rm p}=1/m_{\rm p}$ is the Planck length. Note that this ansatz is
being made in one particular frame, in this case Alice's. Here, $E$ and $p$ are the energy
and momentum as measured by Alice. $|a|$ is the absolute value of an earlier
introduced position vector $a^i$. Since this vector can be arbitrarily chosen, it is 
unclear what its meaning is. However, the ansatz later used for the spread of the wavefunction is that $|a|$ 
is actually the distance the particle travelled, see the later remark in \cite{Smolin:2010xa}, 
after Eq. (65).

That the ansatz Eq.(\ref{ansatz}) should solve the problem pointed out in \cite{Hossenfelder:2009mu}
is puzzling, since it is the same spread of the wave-packet that has been
examined there (see Eq.(13) \cite{Hossenfelder:2009mu} and the paragraph thereafter).
%, which the author of \cite{Smolin:2010xa} somehow fails to mention. 
To be more precise it is the
same spread when one
makes the approximation that the initial width of the photon's wave-function in momentum space is
approximately its energy $\Delta p \approx |p|$. It has been pointed out in \cite{Hossenfelder:2009mu}, that
this corresponds to a quite badly localized particle already. This situation was
examined in \cite{Hossenfelder:2009mu} because it is the most optimistic case when
one tries to make sense of {\sc DSR}, and the most challenging one if one tries to
show it inconsistent. It was probably studied in \cite{Smolin:2010xa} for the same reasons.
Either way, it was shown in \cite{Hossenfelder:2009mu} that even when one considers
such an already badly localized wave-packet the spread cannot entirely hide
the nonlocality. So why then the difference in conclusion?

The
difference in both treatments becomes clear in Eqs. (46,47) \cite{Smolin:2010xa}. There, the
ansatz is transformed into a different restframe, in this case Bob's. This transformation
is done making use of the standard Lorentz-transformations (up to higher order corrections).
However, looking at the ansatz, one sees that $\Delta x_{AliceQ}$ is a product of a
space-time distance and an energy, here the photon's energy. One knows how these both quantities transform, 
the transformations have indeed been used in Eqs. (36,37) \cite{Smolin:2010xa} already. 
The energy transforms under the {\sc DSR} transformations, and the space-time
distance has then to transform accordingly to allow the invariance of the speed of
light in position space.

Consequently, it is the $x$ (or $|a|$ respectively) that is a distance in position
space that transforms according to \cite{Smolin:2010xa} (46,47), whereas with the ansatz 
\cite{Smolin:2010xa} Eqs.(44,45) the postulated uncertainty obtains an additional
factor from the red/blue-shift of the energy appearing therein. Even to leading
order in $l_{\rm p}$ this is not consistent with the transformation behavior used
in \cite{Smolin:2010xa} (46,47). One obtains instead the transformation behavior for
the spread used in \cite{Hossenfelder:2009mu}. It has been shown there that this
is not sufficient to hide the nonlocality in all restframes (see Eqs. (14) - (18)
\cite{Hossenfelder:2009mu}). 

It is on the other hand of course possible to enforce the transformation behavior
postulated in \cite{Smolin:2010xa} (46,47). But then the ansatz Eq. (\ref{ansatz}) is not observer-independent 
in that it would not preserve its form by a change of reference frame. This
ansatz for the uncertainty is, not coincidentally, identical to the difference in arrival times for
photons that had been predicted which is cause of the problem. Giving up its 
observer-independence would mean that the propagation of the photons is either not, as 
claimed by DSR, observer-independent. Or the ansatz does not agree with the delay
in all restframes, which is also not observer-independent. 

We thus come to conclude that one can either indeed hide the nonlocality with
the ansatz proposed in \cite{Smolin:2010xa} but this spoils observer-independence. Or
one preserves observer-independence, and then one recovers the conclusion
previously drawn in \cite{Hossenfelder:2009mu}, that the nonlocality cannot
be hidden in all restframes. This is a consequence of the funny transformation
behavior of the wave-packet's spread that is forced upon us by observer-independence of
the modified dispersion relation.
The problem arises from this transformation being different from that of 
space-time distances defined by other means (the intersection of worldlines), 
and there is always a restframe in which this inconsistency becomes observable. 
To quantify the problem properly, at least three worldlines are necessary to obtain sufficient
intersections.

It is interesting at this point to have a look at another recent paper \cite{Jacob:2010vr} by
Jacob {\it et al}. The authors find in this paper the correct transformation
of the time-delay from one restframe into the other. For the case $n=2$ considered in
\cite{Hossenfelder:2009mu} they find (see Eq. (30) \cite{Jacob:2010vr})
\beqn
\Delta t'_{II} = \frac{1+v}{1-v} \Delta t_{I}~.
\eeqn
This is indeed the same as Eq.(9) \cite{Hossenfelder:2009mu}.

In  \cite{Jacob:2010vr} it is then moreover correctly concluded that this ``poses an
immediate challenge for the consistency of this scenario.'' This inconsistency is
 interpreted to arise from some ``fuzziness'' though it is acknowledged later
that the arising nonlocality is of ``possibly sizeable distan[ce]'' and requires
``a rather drastic change in the description of spacetime.'' The problem
 pointed out in \cite{Hossenfelder:2009mu} is dismissed (see footnote 7 \cite{Jacob:2010vr}) with 
the argument that
\cite{Hossenfelder:2009mu} allegedly assumed that a ``novel geometric description of spacetime could at
best affect the structure of a spacetime point only locally, in a neighborhood
of size $1/E_{QG}$'' (i.e. somewhere close to the Planck scale). In fact, no such
assumption has been made in \cite{Hossenfelder:2009mu}. The bound derived
in  \cite{Hossenfelder:2009mu} is based on allowing a change in spacetime structure
below the limits that the considered interaction is testing known physics. One can alternatively 
understand the calculation in \cite{Hossenfelder:2009mu} the other way round: the 
necessary ``rather drastic''
modification of the spacetime picture to reconcile the inconsistencies in the
{\sc DSR} spacetime picture had to be nonlocal on the km scale at least.  

Both \cite{Jacob:2010vr} as well as \cite{Smolin:2010xa} thus recover the same 
problem pointed out in \cite{Hossenfelder:2009mu}.

\section{And else}

While we are at it, let us have look at the rest of \cite{Smolin:2010xa}. It is comforting,
though not particularly surprising, that the spread of the wave-packet found
in Eq. (65)  \cite{Smolin:2010xa} is the same as that used in \cite{Hossenfelder:2009mu}, since
this a just consequence of the dispersion-relation having a first order
contribution in $E/m_{\rm p}$ together with using a linear superposition.
It remains unclear why the author of \cite{Smolin:2010xa} states that the
problem pointed out in \cite{Hossenfelder:2009mu} is ``not surprising,
because quantum effects are being treated inconsistently'' and then attempts
to address the problem by examining the same spread of the wave-function
already considered in  \cite{Hossenfelder:2009mu}.
It is however good to see and very welcome indeed that this derivation
has been done in a straight-forward 
way starting from the $\kappa$-Poincar\'e algebra. 

The way this
was achieved is to make the novel observation that the non-commutative space-time coordinates $x,t$
in the $\kappa$-Poincar\'e algebra do not make a good set of observables
for quantum mechanics. Instead, it was argued in \cite{Smolin:2010xa} that one needs to
introduce a new time coordinate $T$, see Eq.(8) \cite{Smolin:2010xa},
\beqn
T = t + \frac{x^i p_i}{m_p} e^{-E/m_{\rm p}}~,\label{Tt}
\eeqn
that commutes with $x$. Then, $x$ and $T$ form a complete set of commuting
observables that can be used to describe the propagation in position space. 
One then needs to find the transformations from the $E,p$-basis in momentum
space to the $x,T$-basis in position space. 

It is worthwhile to
note that if one introduces the $T$ already in Eqs.(17), (18), (19) \cite{Smolin:2010xa}, i.e. 
before making the transformation from $E,k$ to $E,p$, then the commutator algebra is simply
the standard one, with $[T,E]=-{\rm i}, [x,k]= {\rm i}$ and all other brackets vanishing. 
One should not be fooled to believe that this means the
physics is the usual too. The physics is contained in the evolution equation,
given by the Hamiltonian (constraint), or the Casimir operator respectively, which in
these $E,k$ coordinates, for massless particles, takes the form
\beqn
k^2 = m_{\rm p}^2\left(1- e^{- E/m_{\rm p}} \right)^2  \label{hamdef}~.
\eeqn
The evolution is thus not identical to the usual one. There are two points we should
take away from here. One is that the transformation Eq.(\ref{Tt}) was constructed
with the intent to change the commutation relations and thus does not constitute
a canonical transformation\footnote{$T$ is in fact not uniquely defined by
the requirement that it brings the algebra in a particular form, but only up to
a canonical transformation. That freedom in the definition however does not
affect the following conclusions about the speed of massless particles.}. 
The second is that either way the physics is contained in
the evolution equation, not in the commutator algebra. 

There is another way than the one taken in \cite{Smolin:2010xa} to arrive at an 
expression for the speed of light with this ansatz which is as follows.
From the commutator relation Eq. (24) \cite{Smolin:2010xa} one obtains
for the momentum operator in position space
\beqn
\hat p = - {\rm i} \partial_x e^{ {\rm i} \partial_T /m_{\rm p}} ~. \label{phat}
\eeqn 
To see this, first note that in position space ($\hat T = T,~ \hat x = x$) since $[T,E] = -{\rm i}$  one has $\hat E = {\rm i} \partial_T$ as
usual. We can then compute $x \hat p (\cdot ) - \hat p x (\cdot)$ with Eq. (\ref{phat}) which yields
\beqn
{\rm i} \partial_x e^{{\rm i} \partial_T /m_{\rm p}} x (\cdot) - x {\rm i} \partial_x e^{ {\rm i} \partial_T /m_{\rm p}} (\cdot) &=& \nonumber\\
{\rm i} e^{{\rm i} \partial_T /m_{\rm p}} (\cdot) \partial_x  x  &=& {\rm i}  e^{{\rm i} \partial_T /m_{\rm p}} (\cdot ) = {\rm i}  e^{\hat E/m_{\rm p}} (\cdot )   \quad.
\eeqn
We have used here that the $x$ and $T$ coordinates commute, since they were constructed this way.
Eq.(\ref{phat}) thus fulfils the right commutator relation. 
%One sees the same way that the
%expression Eq.(66) in \cite{Smolin:2010xa} doesn't. 
An easier way to arrive at $\hat p$ is
to use that $k$ and $x$ fulfill the standard commutation relations, thus $k = - {\rm i} \partial_x$,
and then one uses the definition $p = k \exp(E/m_{\rm p})$ to obtain $\hat p$ in Eq. (\ref{phat}). 
%With Eq.(66) wrong, Eqs.(57), (58), (63), (67) and (68) in \cite{Smolin:2010xa} are also wrong.
%In fact, it would have been easier to reverse the derivation and start with obtaining the
%momentum/energy operator in position space from the commutation relations, and then
%deriving the scalar product $\langle x,T | E,p \rangle$ as an eigenfunction in the following way. 

With this momentum operator, the wave-equation takes the form
 \beqn
m_{\rm p}^2 \left( 1- e^{- {\rm i} \partial_T/m_{\rm p}} \right)^2  + \partial^2_x = 0 \quad.
\eeqn
The solutions to this equations are proportional to
\beqn
\exp(i(xk - ET)) \quad,
\eeqn
with 
\beqn 
k = \pm m_{\rm p}\left(1- e^{- E/m_{\rm p}} \right) \quad,
\eeqn
and the phase velocity is $E/k$ with the above constraint thus
\beqn
c_{p}(E) = \frac{E}{m_{\rm p}} \frac{1}{1- e^{- E/m_{\rm p}}} \quad,
\eeqn
while the group velocity is
\beqn
c_{g}(E) = \frac{1}{dk/dE} = e^{E/m_{\rm p}} ~,
\eeqn
which is in agreement with the later derivation of the group-velocity in Eqs. (71) \cite{Smolin:2010xa} ff \footnote{This calculation has
been corrected in the updated version of \cite{Smolin:2010xa}.}.  
%This mistake in the calculation
%of $\hat p$ does however not matter for the first order expansion in Eq. (70) \cite{Smolin:2010xa}, 
%which is also why it does not really matter for the spread of the wave-packet in Eqs.(64) and (65) \cite{Smolin:2010xa}. 
%In fact, the whole $\kappa$-Poincar\'e prelude is irrelevant for the propagation of
%the wave-packet. 

Recalling what we noted above, we should not be surprised that the result obtained in \cite{Smolin:2010xa} differs
from the result in \cite{Daszkiewicz:2003yr} since the change of the time-coordinate is not a canonical transformation,
and thus the dynamics imposed is physically different in both cases.

Finally, let us look at one of the closing remarks in \cite{Smolin:2010xa}, since it addresses
the ``worst thing that could happen,'' that is 	``if the paradoxes of the classical theory [...] are not
resolved in all cases.'' Since we explained above the worst thing is happening, a comment seems in order.
In a nutshell, the argument offered is that the nonlocal interactions required by {\sc DSR} might not
have been observed yet since the location of the particles would have to be very exactly chosen in order
for a nonlocal interaction to take place. If we think again about the one point that splits 
up into three points by a change of reference frame, 
it might seem that observer-independence would merely require particles on exactly two of these three points to 
interact with the same probability as the particles in the one point in the original frame. This is
indeed an excellent objection, and one that has not been addressed in \cite{Hossenfelder:2009mu}. 

However, the argument neglects to take into account that the splitting of the point does not
only depend on the phase-space of each individual particle, it does also depend on the distance
the particles have travelled. To see this, recall we noted earlier that what causes the splitting
of the point to become macroscopically large is what also causes the time-delay of the highly-energetic
photon to be observable in the first place: the long distance travelled. Thus, the location of
the three points that a change of reference frame creates out of the original one point also depend
on this distance. In the more general case it would depend on the distances
of all the particles involved in defining the point. The interaction probability had to be the same
for all of these points created by all of these distances. Unless there is a not obvious degeneracy in
these locations, the interaction points should cover a subset of a 2-dimensional surface. The
interaction probability would have to be the same for any two points in that subset, which 
would vastly increase in-medium (or close-to medium) interactions. In any case, this option might
deserve further investigation.
  
\section{Conclusion}

The ansatz proposed in \cite{Smolin:2010xa} for the spread of the wave-packet
does not solve the problem discussed in \cite{Hossenfelder:2009mu}. \cite{Jacob:2010vr} 
confirms the same problem, but the authors stop short from quantifying it and 
instead interpret the inconsistency as the arising necessity for quantum ``fuzziness.''  

\section*{Acknowledgements}
I thank Lee Smolin and Stefan Scherer for helpful comments. 
This work was supported by the Queen of Hearts under grant 299792458.

\end{document}